\documentclass{article}
 \newcommand{\beq}{\begin{equation}}
\newcommand{\eeq}{\end{equation}} 
\newcommand{\bea}{\begin{eqnarray}}
\newcommand{\eea}{\end{eqnarray}} \newcommand{\nn}{\nonumber}

 \newcommand{\qm}{quantum
mechanics} 
\newcommand{\ca}{$C^*$-algebra} 
 \newcommand{\rep}{representation}

 \newcommand{\ovl}{\overline}
 
\newcommand{\raw}{\rightarrow}

\newcommand{\ot}{\otimes} 
\newcommand{\la}{\langle} \newcommand{\ra}{\rangle}

\newcommand{\x}{\times} 
 
\newcommand{\half}{\mbox{\footnotesize $\frac{1}{2}$}}

\newcommand{\er}{\eqref}
\newcommand{\dl}{\delta} \newcommand{\Dl}{\Delta}
\newcommand{\ep}{\epsilon} \newcommand{\varep}{\varepsilon}

\newcommand{\lm}{\lambda} 
\newcommand{\rh}{\rho} \newcommand{\sg}{\sigma}
  
 \newcommand{\phv}{\varphi}
\newcommand{\ch}{\chi} \newcommand{\ps}{\psi} 
\newcommand{\om}{\omega} \newcommand{\Om}{\Omega}
\newcommand{\GS}{\mathfrak{S}}

\newcommand{\CA}{{\mathcal A}} \newcommand{\CB}{{\mathcal B}}

 \newcommand{\CS}{{\mathcal S}}
 \newcommand{\CP}{{\mathcal P}}

\newcommand{\C}{{\mathbb C}} 
 
\newcommand{\N}{{\mathbb N}} \newcommand{\R}{{\mathbb R}}
 
  \makeatletter
\newskip\tempskip \def\endproof{{\parfillskip24\p@ plus\@ne
fil\@@par}\tempskip\prevdepth
\ifdim\lastskip=\z@\tempskip\z@\else\vskip-\lastskip
\ifdim\tempskip>4\p@ \tempskip.5\tempskip \else \tempskip\z@\fi\fi
\nobreak\vskip-\baselineskip\vskip-\tempskip\noindent\hbox
to\hsize{\hfill
$\blacksquare$}\par\vskip\tempskip\vskip\abovedisplayskip\@doendpe}
\makeatother \makeatletter
\newskip\tempskip \def\endiproof{{\parfillskip24\p@ plus\@ne
fil\@@par}\tempskip\prevdepth
\ifdim\lastskip=\z@\tempskip\z@\else\vskip-\lastskip
\ifdim\tempskip>4\p@ \tempskip.5\tempskip \else \tempskip\z@\fi\fi
\nobreak\vskip-\baselineskip\vskip-\tempskip\noindent\hbox
to\hsize{\hfill
$\Box$}\par\vskip\tempskip\vskip\abovedisplayskip\@doendpe}
\makeatother \newcommand{\enp}{\endproof}

\topmargin = - 0.5 cm
\usepackage{amsmath}
\usepackage{amssymb}
\newtheorem{Definition}{Definition}

\newtheorem{Theorem}{Theorem}

\newcommand{\ut}{\underline{2}}
\newcommand{\lni}{\lim_{N\raw\infty}}
\begin{document}
\setlength{\baselineskip}{1\baselineskip}
\thispagestyle{empty}
\title{Macroscopic observables and the Born rule \\ I. Long run frequencies}
\author{\textbf{N.P. Landsman} \\ \mbox{} \hfill \\
Institute for Mathematics, Astrophysics, and Particle Physics\\
Faculty of Science\\
Radboud University Nijmegen\\
Toernooiveld 1, 6525 ED NIJMEGEN\\
THE NETHERLANDS\\
\mbox{} \hfill \\
email \texttt{landsman@math.ru.nl}}
\date{\today}
\maketitle
\smallskip
\begin{center}{\it
Dedicated to the memory of Bernd Kuckert (1968--2008)}
\end{center}
\bigskip
\begin{abstract}
We  clarify the role of the Born rule in the Copenhagen Interpretation of quantum mechanics by deriving it from Bohr's doctrine of classical concepts, translated  into the following mathematical statement: a quantum system 
described by a noncommutative \ca\  of observables 
is empirically accessible only through associated commutative \ca s. 
The Born probabilities emerge as the relative frequencies of outcomes in long runs of measurements on a quantum system; it is {\it not} necessary to adopt the frequency interpretation of single-case probabilities (which will be the subject of a sequel paper).
Our derivation of the Born rule uses ideas from a program begun by Finkelstein (1965) and Hartle (1968), intending to remove the Born rule as a separate postulate of quantum mechanics. Mathematically speaking, our approach refines previous elaborations of this program - notably 
the one due to  Farhi, Goldstone, and Gutmann (1989) as completed by Van Wesep (2006) - in replacing infinite tensor products of Hilbert spaces by continuous fields of \ca s. 
In combination with our interpretational context, this technical improvement circumvents valid criticisms that earlier derivations of the Born rule have provoked, especially to the effect that such derivations were mathematically flawed as well as circular. Furthermore, instead of relying on the controversial eigenvector-eigenvalue link in \emph{quantum theory}, our derivation just assumes that 
 pure states in \emph{classical physics} have the usual interpretation as truthmakers that
assign sharp values to observables.
\end{abstract}\newpage
\section{Introduction}
In its simplest formulation, the Born rule says that if $A$ is some quantum-mechanical observable with nondegenerate discrete spectrum $\sg(A)$, then  the probability
$P_{\ps}(A=\lm_i)$
 that a measurement of $A$
in a state $|\psi\ra$ yields the result $\lm_i\in\sg(A)$ is given by 
\beq
P_{\ps}(A=\lm_i)=
|\la e_i,\psi\ra|^2,\eeq where $|e_i\ra$ is a normalized eigenvector of $A$ with eigenvector $\lm_i$.  In other words, if $|\psi\ra=\sum_i c_i|e_i\ra$ with $\sum_i |c_i|^2=1$, then $P_{\ps}(A=\lm_i)= |c_i|^2$.
 More generally, if $A$ is a self-adjoint operator on a Hilbert space $H$ with associated spectral measure
$\Delta\mapsto E(\Dl)$, then the probability $P_{\ps}(A\in \Dl)$ that the proposition $A\in\Dl$ comes out to be  true if $A$ is measured in a state $|\psi\ra$ equals 
\beq P_{\ps}(A\in \Dl)=
\la \ps| E(\Dl)|\ps\ra.\label{BornvN}\eeq

The Born rule provides the key  link between the mathematical formalism of quantum physics and experiment, and as such is responsible  for  most predictions of quantum theory. In the history and philosophy of science,  the Born rule (on a par with the Heisenberg uncertainty relations) is often seen as a turning point where indeterminism entered fundamental physics.\footnote{ 
The Born rule was first stated by Max Born in the context of scattering theory \cite{born}, following a slightly earlier paper in which he famously omitted the absolute value squared signs (though he corrected this in a footnote added in proof). The well-known application to the position operator  is due to Pauli \cite{pauli}. 
The general formulation \er{BornvN} is due to von Neumann \cite[\S {\sc iii}.1]{vN}. See \cite{20}
for a detailed reconstruction of the historical origin of the Born rule within the context of quantum mechanics, 
as well as \cite{22} for a briefer historical  treatment in the more general setting of the emergence of modern 
probability theory and probabilistic thinking.} Of course, classical physics is full of random phenomena as well. But in all known cases, their apparent random character may be retraced to ignorance about the initial state or about microscopic degrees of freedom or time scales; see, e.g.,  \cite{Engel} and \cite{22}. 
In contrast, the type of randomness to which  quantum mechanics gives rise via the Born rule is generally felt to be `irreducible'  (in the sense of not being reducible to ignorance, not even about the Laws of Nature).\footnote{Notable exceptions are Einstein \cite[pp.\ 129--130]{EB} and 't Hooft \cite{hooft}.}

Even the assumption that quantum mechanics is a correct and fundamental theory by no means implies that 
 this feeling is correct. Indeed,  although among rival interpretations the Copenhagen Interpretation is the one that arguably puts most emphasis on both the fundamental and the probabilistic character of quantum theory,  
 a mature work by one of its founders actually contains the following passage:
\begin{quote} `One may call these uncertainties objective, in that they are simply a consequence of the fact that we describe the experiment in terms of classical physics; they do not depend in detail on the observer. One may call them subjective, in that they reflect our incomplete knowledge of the world.' (Heisenberg, \cite[pp.\ 53--54]{heis}.)
\end{quote}
This claim is in tune with one of the two main principles of the Copenhagen Interpretation,\footnote{The other
one, the Principle of Complementarity, plays no role in this paper.}
 namely Bohr's
{\it doctrine of classical concepts}. A mature and well-known expression of this doctrine is as follows:
 \begin{quote}
 `However far the phenomena transcend the scope of classical physical explanation, the account of all evidence must be expressed in classical terms. (\ldots) The argument is simply that by the word {\it experiment} we refer to a situation where we can tell others what we have done and what we have learned and that, therefore, the account of the experimental arrangements and of the results of the observations must be expressed in unambiguous language with suitable application of the terminology of classical physics.'
  (Bohr, \cite[p.\ 209]{bohr}.)
 \end{quote}
Elsewhere, Bohr time and again stresses that measurement devices must be described classically
 `if these are to serve their purpose'. We take this to mean that, although such devices are \emph{ontologically quantum-mechanical} by nature,  they become a tool (in fact, the only tool) for the description of quantum phenomena as soon as they are epistemically treated \emph{as if they were classical}. Thus the so-called {\it Heisenberg cut}, i.e.\ the borderline between the part of the world that is described classically and the part that is described 
 quantum-mechanically, is epistemic or (inter)subjective in nature and hence movable; see also \cite{24,SC}.
In our opinion, this ideology provides an attractive qualitative basis for the understanding of randomness in Nature, for it preserves the fundamental difference between random phenomena in classical and in quantum physics (the given explanation of quantum probabilities as arising from the classical description of some part of the world would not make any sense if applied to classical probabilities),  while discarding the notion of strictly `irreducible' randomness (which is only defined through negation and quite possibly makes no philosophical sense at all). 

However, little (if any) work has been done in relating this ideology to the Born rule, which in the Copenhagen Interpretation simply seems to be taken for granted as a mathematical recipe that requires no explanation. It is the purpose of the present paper to fill this gap: as we shall see, the Born rule can actually be derived from a particular instance of Bohr's doctrine of classical concepts, provided one identifies the Born probabilities with the relative frequencies of  outcomes in long runs of measurements on a quantum system.\footnote{With this limited goal it is not even necessary to \emph{mention} single-case probabilities, let alone \emph{interpret} them; doing so requires a far deeper analysis, which will be the subject of a sequel paper, based on  \cite{HLS1}.} As always, the mathematical implementation of Bohr's philosophical ideas is ambiguous; as far as his doctrine of classical concepts is concerned,
we read it as saying that  a quantum system described by a noncommutative algebra $A$ of observables  is empirically accessible only through commutative algebras associated with $A$.\footnote{Apart from leading to the Born rule, this reading also gives rise to a very pretty description of complementarity through the mathematical framework of topos theory; see \cite{HLS1}. Cf.\  Scheibe \cite{24} and Howard  \cite{How} for different readings of the doctrine of classical concepts.} 
For convenience, and in line with the modern mathematical description of quantum theory 
\cite{T4,BR2,Haag,Sewell,book,handbook}, we assume that these algebras are in fact (unital) \ca s. The simplest kind of commutative algebras associated with $A$ are its (unital) commutative
 $C^*$-subalgebras; in this paper we need a more subtle limiting procedure to `extract' a
 commutative \ca\ of macroscopic observables.

Our derivation of the Born rule relies on certain ideas that were originally proposed by  Finkelstein \cite{16} and Hartle \cite{19}, whose work was continued by Ochs \cite{ochs},
Bugajski and Motyka \cite{BM}, Farhi, Goldstone and Gutmann \cite{14}, and Van Wesep \cite{vW}.\footnote{Some of these papers were not quite written in support of the  Copenhagen Interpretation but rather against it, usually defending the Everettian stance.}
We review this development in Section \ref{sec2}, either incorporating or circumventing  critique of  the papers just listed  that has been issued 
  by a number of authors, including
 Cassinello and  S\'{a}nchez-G\'{o}mez \cite{11} and Caves and Schack \cite{12}. Their critique has been partly of a mathematical and partly of a conceptual nature, but in our opinion one of the most devastating arguments against the program in question, namely its reliance on the so-called eigenstate-eigenvalue link, has not been made before.   
 
We will show how the program of deriving the Born rule from `first principles' can nonetheless be carried out if it is underwritten by  Bohr's doctrine of classical concepts (in its reading mentioned above). The mathematical formalism needed to accomplish this starts from the modern algebraic approach to the quantum theory of large systems \cite{T4,BR2,Haag,Hepp,MS,Sewell}, which, however, we need to reformulate in order to incorporate  Bohr's doctrine in an optimal way. This reformulation is based on the unified picture provided by continuous fields of \ca s \cite{Dix,KW} in the description of the classical limit of quantum mechanics. This limit actually has (at least)  two guises, namely the limit $\hbar\raw 0$ of Planck's constant going to zero, and the limit $N\raw \infty$ of a system size going to infinity. Both can be brought under the umbrella of continuous fields of \ca s; for  $\hbar\raw 0$ this was done in  \cite{book}, and for  $N\raw \infty$ it was announced in  \cite{handbook}
and will be completed in the present paper,
where essential use is made of ideas of Raggio and Werner \cite{RW} and Duffield and Werner \cite{DW}.  
 In fact, once the appropriate framework has been set up in Section \ref{sec3}, the derivation of the Born rule in Section \ref{sec4} will turn out to be almost trivial.  
 
 This paper is part of a larger research programme, whose goal it is to interpret quantum mechanics entirely in terms of its classical limit. This is meant as a technical implementation of the Copenhagen Interpretation as originally formulated by Bohr and Heisenberg (cf.\ \cite{HowC}), whose goal was expressed quite well by Landau and Lifshitz \cite[p.\ 3]{LL}:
  \begin{quote}
``Thus \qm\ occupies a very unusual place among physical theories: it contains classical mechanics as a limiting case, yet at the same time it requires this limiting case for its own formulation.''
\end{quote} 
\section{The strong law of large numbers in quantum theory} \label{sec2}
Let us first review what has been achieved mathematically in \cite{16,19,ochs,BM,14,vW}.
For simplicity, we restrict ourselves
to the simple situation of repeated measurements on a two-level (or, in current parlance, one-qubit) system, i.e.\ with Hilbert space $\C^2$. 
Suppose we have an observable $A$ 
(i.e.\ a hermitian $2\x 2$ matrix) with eigenvalues $0$ and $1$ and 
corresponding orthogonal eigenstates $|0\ra$ and $|1\ra$.  A long series of measurements of $A$ in a given initial state $|\ps\ra\in\C^2$ (prepared anew for each subsequent measurement) will produce a sequence $x=(x_1,x_2,\ldots)$, where $x_i=0$ or $1$. We idealize a long series of measurements as an infinite one, so that $x\in \underline{2}^{\N}$, with $\underline{2}=\{0,1\}$ and the space of infinite binary sequences is denoted by $\underline{2}^{\N}=\{x:\N\raw\ut\}$.
 We define $p\in [0,1]$ as the Born probability 
\beq p=|\la 1|\ps\ra|^2,\label{defp}\eeq so that $|\la 0|\ps\ra|^2=1-p$.

We first review the classical strong law of large numbers relevant to $\ut^{\N}$, seen as a measure space
with Borel structure generated by the sets 
\beq B_k^{(\ep)}=\{x:\N\raw\ut\mid x_k=\ep\},\label{Bor1}
\eeq
 where
$k\in\N$ and $\ep\in\ut$.
For any $p\in [0,1]$, consider the probability measure $\mu_p$ on $\underline{2}$ defined by $\mu_p(0)=1-p$ and $\mu_p(1)=p$. This defines a probability measure $\mu_p^{\infty}$ on $\underline{2}^\N$ for which
$\mu_p^{\infty}(B_{k}^{(1)})=p$ and $\mu_p^{\infty}(B^{(0)}_k)=1-p$ for all $k$. Let
\beq L_p=\left\{x\in\ut^\N\mid \lim_{N\raw\infty} \frac{1}{N}\sum_{k=1}^N x_k=p\right\}\subset\ut^\N.\label{Lp}\eeq
This is a Borel set.
The  strong law of large numbers states that
\beq \mu_p^{\infty}(L_p)=1.\label{slln} \eeq
A measure theorist will read this as is stands: $L_p$ has measure one with respect to 
$\mu_p^{\infty}$. A probability theorist defines functions $f_k:\ut^\N\raw\ut$
by $f_k(x)=x_k$, notes that the $f_k$ are i.i.d.\ random variables, and says that the sequence 
of functions $(1/N)\sum_{k=1}^N f_k$ on $\ut^\N$ converges pointwise to $p$ 
with probability one (or almost surely) with respect to  $\mu_p^{\infty}$. A physicist 
defines an elementary proposition (or `yes-no question') $\ch_{L_p}$ (i.e.\ the characteristic function of $L_p$) on the `phase space' $\ut^\N$, which is answered by yes in a pure state $x$ if $\lim_{N\raw\infty} \frac{1}{N}\sum_{k=1}^N x_k=p$, and by no otherwise. The probability measure  $\mu_p^{\infty}$ defines a mixed state on $\ut^\N$, and \er{slln} gives the state-proposition pairing in the case at hand as
\beq
\la \mu_p^{\infty}, \ch_{L_p}\ra=1.\label{prob1}\eeq
 If, for a general yes-no question $Q$ and state $\rh$, one 
initially interprets $\la\rh,Q\ra$ as the probability of obtaining a positive answer to $Q$ in the state $\rh$ (or, more generally, interprets $\la\rh,f\ra$ as the expectation value of an observable $f$ in a state $\rh$), then one still has to expand this interpretation by stipulating what notion of probability one is using \cite{17,mellor}. Even if a probability equals one, as in \er{prob1}, one still has to declare whether or not one adopts the so-called {\it Necessity Thesis} \cite{mellor} (stating that probability one implies certainty). These questions cannot be answered by the mathematical formalism.

The papers just cited attempt to extend the strong law of large numbers to the quantum case, and, not always sensitive to the last remark, draw certain conclusion about quantum mechanics from such an extension. A correct way of proceeding at least mathematically 
emerges from a combinination of results in \cite{14,vW}, as follows. Let $(\C^2)^{\ot N}\cong \C^{2N}$ be the $N$-fold tensor product of $\C^2$, and let $(\C^2)_{\ps}^{\ot\infty}$ be the separable component of the infinite tensor product $(\C^2)^{\ot\infty}$ Hilbert space (in the sense of von Neumann \cite{vNC}) of $\C^2$ that contains $|\ps\ra^{\ot\infty}$, where $|\ps\ra\in\C^2$ is a given (unit) vector.\footnote{Apart from the original source \cite{vNC}, this formalism is also explained
in e.g.\ \cite{14} or \cite[\S 6.2]{EK}. The details are not relevant here, as we will replace the use of infinite tensor products of Hilbert spaces by a different formalism later on. In any case, the simplest way to define $(\C^2)_{\ps}^{\ot\infty}$ is to regard it as the Hilbert space of the GNS-representation of the infinite tensor product $\bigotimes^{\N} M_2(\C)$ (cf.\ \cite[\S 11.4]{KR2})
induced by  the vector state $|\ps\ra^{\ot\infty}$;
see Section \ref{sec4} below.} The unit vector  $|\ps\ra^{\ot\infty}\in (\C^2)_{\ps}^{\ot\infty}\subset  (\C^2)^{\ot\infty}$, seen as a state, is the quantum analogue of the probability measure $\mu_p^{\infty}$ in the classical situation just reviewed.
The quantum analogue of the proposition $\ch_{L_p}$ is a projection $\mathcal{P}(L_p)$
on $(\C^2)_{\ps}^{\ot\infty}$, defined as follows. For each $k\in \N$, define a projection
$P^{(1)}_k$ on $(\C^2)_{\ps}^{\ot\infty}$ by $P^{(1)}_k=1\ot\cdots |1\ra\la 1|\ot 1\cdots$, where the projection $|1\ra\la 1|$ on $\C^2$ acts on the $k$'th copy of $\C^2$ in the infinite tensor product and all other entries are unit matrices on $\C^2$. Similarly, $P^{(0)}_k$ is defined by replacing $|1\ra\la 1|$ by $|0\ra\la 0|$. 
The projections $\{P^{(0)}_k,P^{(1)}_k\}_{k\in\N}$ commute, and generate a complete Boolean algebra $\mathfrak{P}_{\ps}$ of projections on $(\C^2)_{\ps}^{\ot\infty}$. Let $\mathfrak{B}_{\ut^\N}$ be the (countably complete) Boolean algebra of Borel sets in $\ut^\N$.
By  Theorem 1 in \cite{vW}, there is a  unique homomorphism 
$\CP:\mathfrak{B}_{\ut^\N}\raw \mathfrak{P}_{\ps}$ of Boolean algebras that satisfies
\begin{eqnarray}
\CP(B_k^{(0)})&=&P^{(0)}_k;\nn\\
\CP(B_k^{(1)})&=&P^{(1)}_k,
\end{eqnarray}
for each $k\in\N$.  The projection  $\mathcal{P}(L_p)$, then,
is what its notation says, i.e.\  the image of the Borel set $L_p\in\mathfrak{B}_{\ut^\N}$
 under $\CP$. Interpreted as a yes-no question, it asks if a given measurement outcome $x$ has mean $p$.

Let $p$ be as in \er{defp}. It is easy to show \cite{vW} that \er{slln} implies
\beq
\mathcal{P}(L_p)|\ps\ra^{\ot\infty}=|\ps\ra^{\ot\infty}.\label{wesep1}\eeq
Regarding the unit vector $|\ps\ra^{\ot\infty}$ as a state $\ps^{\ot\infty}$ (in the algebraic sense)
on any von Neumann algebra of operators on $(\C^2)_{\ps}^{\ot\infty}$ containing $\mathcal{P}(L_p)$, we can rewrite \er{wesep1} in the form of the classical pairing \er{prob1}, i.e.
\beq \la \ps^{\ot\infty},\mathcal{P}(L_p)\ra=1.\label{wesep2}\eeq
Indeed, since $\mathcal{P}(L_p)$ is a projection, eqs.\ \er{wesep1} and \er{wesep2}
are equivalent. 

Furthermore, let us define the {\it frequency operator} $f_N$ 
on $\C^{2N}$ by stipulating that its eigenstates are
$|x_1\ra\cdots |x_N\ra$ (where $x_i=0$ or $1$), with eigenvalues
\beq f_N|x_1\ra\cdots |x_N\ra=\frac{1}{N} \sum_{k=1}^N x_k |x_1\ra\cdots |x_N\ra.\label{deffN}\eeq
In words, $f_N$ is the relative frequency of the entry 1 in the list $(x_1,\ldots, x_N)$.
Clearly, $f_N$ can be extended to an operator $f_N^{\ps}$ 
on  $(\C^2)_{\ps}^{\ot\infty}$ by
\beq f_N^{\ps}=\frac{1}{N} \sum_{k=1}^N P_k^{(1)}.\label{abuse}\eeq
It then follows from \er{wesep1} that
\beq 
\lim_{N\raw\infty}  f_N^{\ps} |\ps\ra^{\ot\infty}=p|\ps\ra^{\ot\infty},\label{wesep3}\eeq
with $p$ given by \er{defp}.
In fact, defining
\beq f_{\infty}^{\ps}=\mathrm{s-}\lim_{N\raw\infty}  f^{\ps}_N,\label{flim}\eeq
where $\mathrm{s-}\lim$ denotes the limit in the strong operator topology on  $(\C^2)_{\ps}^{\ot\infty}$,
 it can even be shown that
\beq 
f_{\infty}^{\ps} =p\cdot 1, \label{fop}\eeq where $1$ is the unit operator on $(\C^2)_{\ps}^{\ot\infty}$.

Results of this type can be derived quite easily from the modern algebraic approach to the quantum theory of large systems \cite{T4,BR2,Haag,handbook}; see below. For the moment,
we  discuss the interpretation of \er{fop} and especially of its corollary \er{wesep3}. 

Authors of papers like \cite{16,19,14,vW} argue that, in view of \er{defp} and the definition \er{abuse} of $f^{\ps}_N$, eq.\ \er{wesep3} provides a derivation of  the Born rule from the so-called \emph{eigenstate-eigenvalue link}. This terminology, which sounds like a tautology in mathematics,  is often used in the philosophy of physics; see, e.g. \cite{Bub,Dickson}. The link in question is the postulate that if $A$ is an observable and $|\ps\ra$ is an eigenstate of $A$ with eigenvalue $\lm$, then a measurement
of $A$ in the state  $|\ps\ra$ yields the result $\lm$ with certainty.\footnote{Philosophical realists adhering to the eigenstate-eigenvalue link would simply say that $A$ \emph{has} the value $\lm$ in the eigenstate  $|\ps\ra$, but precisely among realists it has become fashionable to deny the eigenstate-eigenvalue link for the reasons mentioned in the main text \cite{Bub,Dickson}.
For example, in Bohmian mechanics position always has a sharp value, whereas in the modal interpretation of quantum mechanics the link is dropped in a more flexible way.
}
The eigenstate-eigenvalue link is, of course, a special 
case of the Born rule, but the whole point of the exercise is to derive the Born rule from the eigenstate-eigenvalue link, rather than the other way round.

To assess the claim that  \er{wesep3} provides a derivation of  the Born rule from the eigenstate-eigenvalue link, we make three points.\footnote{We leave it to the reader to assess
the more far-reaching claim in  \cite{vW} that \er{wesep3} ``nullifies any remaining objection to the many-worlds view''. Given their recent attempt to derive the Born rule in a completely different way  \cite{23,26}, this claim is apparently not even supported by adherents of the many-worlds interpretation.}
\begin{enumerate}
\item Applying  the  eigenstate-eigenvalue link to conclude from \er{wesep3} that
$f_{\infty}^{\ps}$ has a sharp value $p$ in the state $|\ps\ra^{\ot\infty}$, is inconsistent with measuring
$f_{\infty}^{\ps}$ by measuring each of its components $P^{(1)}_k$ in \er{abuse} separately. For each such measurement will disturb the state; neither $|\ps\ra^{\ot\infty}$ nor any $|\ps\ra^{\ot N}$ is an eigenstate
of \emph{any} $P^{(1)}_k$, not to mention all of them.\footnote{In fact, any component of the complete von Neumann tensor product
$(\C^2)^{\ot\infty}$ containing at least one simultaneous eigenstate of all $P^{(1)}_k$ is orthogonal in its entirety to $(\C^2)_{\ps}^{\ot\infty}$.}
Hence $f$ is to be measured directly. Although according to \cite{14} this can be done in some cases, it precludes any inference of single-case Born probabilities from
 \er{wesep3}.
\item Even if a measurement of $f_{\infty}^{\ps}$ were to take place by computing the limit \er{flim}
from an infinite list $x$ of single-case measurements, interpreting Born probabilities as
limiting frequencies would face all the usual objections to the frequency interpretation of probability \cite{15,17,18,mellor}.
\item The eigenstate-eigenvalue link is the source of the measurement problem in quantum mechanics and hence is held to be unsound by most contemporary specialists in the foundations of quantum mechanics (see \cite{Bub,Dickson} and references therein). 
Indeed, the eigenstate-eigenvalue link  cannot be found in the writings of Bohr and Heisenberg; 
it was first postulated by Dirac \cite{Dir}.
\end{enumerate}
 The first two points were also made in \cite{11,12}, but despite our adding the third objection, we hesitate in following the authors of these papers in concluding that the program of deriving the Born probabilities from properties of the frequency operator is ``flawed at every step'' \cite{12}. Indeed, by changing both the conceptual and the mathematical setting we will see that each of these objections can be met:
 \begin{enumerate}
\item Changing the definition of the frequency operator $f_{\infty}^{\ps}$  from a strong operator limit 
on a Hilbert space (which even depends on the state $|\ps\ra$) to an element $f_{\infty}$ of a 
\emph{commutative} (i.e.\ classical) \ca\  of macroscopic observables (which is independent of  $|\ps\ra$) `stabilizes' $f$ against perturbations. Thus, without jeopardizing our derivation and interpretation of the Born rule, the  frequency operator can be measured either directly (as suggested in \cite{14}), or in terms of repeated measurements of the underlying observable $A$ in the state $|\psi\ra$. The latter procedure 
 determines the possible values (0 or 1) of each $P_k^{(1)}$ for $k=1,\ldots, N<\infty$, upon which one takes the limit $N\raw\infty$. This seems to correspond to experimental practice.
\item The second objection is obviated if one simply interprets the possible values of the frequency operator $f_{\infty}$ according to  its definition,  i.e.\ as limiting frequencies of either a single experiment on a large number of sites or 
a long run of individual experiments on single sites. In particular, one should refrain from making any statement about single-case probabilities.
On this view, the Born rule simply says \emph{nothing} about individual experiments on single sites.\footnote{Except in an empty way, as in Popper's so-called propensity interpretation of probability \cite{17,mellor}.}
\item  Instead of relying on the controversial eigenvector-eigenvalue link in \emph{quantum theory}, our derivation will just assume that pure states in \emph{classical physics} have the usual interpretation as  `truthmakers' that assign sharp values to observables.
\end{enumerate}
\section{Large quantum systems and the Born rule}\label{sec3}
\subsection{Continuous field of \ca s}
In experimental physics, theoretical predictions based on the Born rule are typically checked by performing $N$ identical experiments on a given quantum system in a given state $|\psi\ra$, where $N$ is large. This situation is idealized by taking the limit $N\raw\infty$. We describe this limit in a way that reorganizes the well-known algebraic description of infinite quantum systems  
by quasilocal \ca s  \cite{T4,BR2,Haag} and macroscopic observables \cite{Hepp,MS,RW,DW,Sewell}
into our preferred tool in the mathematical analysis of classical behaviour in quantum theory
\cite{book,handbook}, namely continuous fields of \ca s. For the reader's convenience we recall  the latter notion, replacing the original definition of Dixmier  \cite{Dix} by the equivalent formulation of Kirchberg and S. Wassermann \cite{KW}. By a morphism we mean a $\mbox{}^*$-homomorphism.
\begin{Definition}\label{defcfca}
  A \textbf{continuous field of $C^*$-algebras} over a
locally compact Hausdorff space $X$   consists of a
 \ca\ $\CA$, a collection of \ca s $\{\CA_{x}\}_{{x}\in X}$, and a surjective morphism $\phv_{x}:\CA\raw\CA_{x}$ for each $x\in X$, such that:
\begin{enumerate}
\item
The function ${x}\mapsto  \|\phv_{x}(A)\|_{x}$ is in $C_0(X)$ for each $A\in\CA$ (where $\|\cdot\|_{x}$ is the norm in $\CA_{x}$).
\item
The norm of $A\in\CA$ is $\| A\|=\sup_{{x}\in X}\|\phv_{x}(A)\|$.
\item The \ca\
$\CA$ is a $C_0(X)$ module in the sense that
for any $f\in C_0(X)$ and $A\in\CA$ there is an element $fA\in\CA$ for
which $\phv_{x}(fA)=f({x})\phv_{x}(A)$ for all ${x}\in X$.
\end{enumerate}
A \textbf{continuous section}
of the field is a map $x\mapsto A_x \in \CA_{x}$
 for which there is an $A\in \CA$ such that $A_{x}=\phv_{x}(A)$ for all ${x}\in X$.
 \end{Definition}
 It follows that
the \ca\ $\CA$ may actually be identified with the space of continuous sections of the field: if we do so, the morphism $\phv_{x}$ is just the evaluation map at $x$. The general idea is that  the  family $(\CA_x)_{x\in X}$ of \ca s is glued together by specifying a topology on the bundle 
$\coprod_{x\in X}\CA_x$ (disjoint union). This topology is defined indirectly via the specification of the space of continuous sections of the bundle (cf.\ the Serre--Swan Theorem for vector bundles). The third condition makes $\CA$ a $C_0(X)$-module in the sense that there exists 
a nondegenerate morphism from $C_0(X)$ to the center of the multiplier algebra of $\CA$. 

This seemingly technical definition turns out to provide an attractive framework for the study of the classical limit of quantum mechanics. In the scenario $\hbar\raw 0$, the parameter space $X$ is typically $X=[0,1]$, and $\CA_0$ is the commutative \ca\ of $C_0$-functions on some classical phase space. For each $\hbar>0$, one then constructs $\CA_{\hbar}$ as the algebra of quantum observables for varying $\hbar$ (it may or may not be the case that the $\CA_{\hbar}$ are isomorphic for different values of $\hbar$). Continuous sections of the field then describe quantization and the classical limit of observables at one go \cite{book}. More generally, the classical theory is `glued' to the corresponding quantum theories via the continuous field structure. 
 \subsection{Macroscopic and quasilocal observables}\label{MO}
To describe large quantum systems and their possible classical behaviour, we use
the one-point compactification $X=\dot{\N}$. This is homeomorphic to 
 $\{0\}\cup 1/\N\subset \R$ in the relative topology borrowed from $\R$, viz.\ 
under the map $n\mapsto 1/n$ and $\infty\mapsto 0$ (where $\infty$ is the compactification point added to $\N$).

To derive the Born rule, we need  $N$  copies of a single quantum system with unital algebra of observables $\CA_1$ (e.g., $\CA_1=M_2(\C)$ as above).
 From the single
\ca\ $\CA_1$, we are going to construct two quite different continuous fields of \ca s over $\dot{\N}$, called $\CA^{\mathrm (c)}$ and $\CA^{\mathrm (q)}$. 
These fields coincide as far as their fibers above $N\in\N$ are concerned, which are given by
\beq 
\CA_{N}^{\mathrm (c)}=\CA_{N}^{\mathrm (q)}=
 \CA_1^{\ot N}.\label{eerst} \eeq
 Here $\ot$ is the  {\it spatial} tensor product; see, e.g., \cite[Ch.\ 11]{KR2}.
  However, the two fields differ in their respective fibers above the limit point $\infty$, given by
 \begin{eqnarray}
\CA_{\infty}^{\mathrm (c)} &=& C(\CS(\CA_1)); \label{fiberc}\\
\CA_{\infty}^{\mathrm (q)} &=& \underline{\lim}_{N}  \CA_1^{\ot N}.
 \label{fiberq}
\end{eqnarray}
 Here  $\CS(\CA_1)$ is the state space of $\CA_1$ (equipped with the weak$\mbox{}^*$-topology),\footnote{For example, the state space of  $\CA_1=\CS(M_2(\C))$ is  isomorphic as a compact convex set to the three-ball $B^3=\{(x,y,z)\in\R^3\mid x^2+y^2+z^2\leq 1\}$:
  describing a state as a density matrix $\rh$ on $\C^2$, the corresponding point $(x,y,z)\in B^3$ is given by the well-known parametrization $
\rh(x,y,z)=\half   \left(\begin{array}{cc} 1+z & x-iy \\ x+iy & 1-z \end{array}\right)
$.}  and the \ca\ in the right-hand side of \er{fiberq} is the inductive limit 
  with respect to the inclusion maps $\CA_1^{\ot N}\hookrightarrow\CA_1^{N+1}$ given by $A_N\mapsto A_N\ot 1$ (see below for an explicit description).\footnote{One often writes
  $\overline{\cup_{N\in\N} \CA_N}$ for the inductive limit $\underline{\lim}_{N} \CA_N$, where the bar denotes norm completion. In the notation of \cite[\S 11.4]{KR2},  our 
  $\underline{\lim}_{N}  \CA_1^{\ot N}$ corresponds to $\bigotimes_{a\in\mathbb{A}}\mathfrak{A}_a$ with $\mathbb{A}=\N$ and $\mathfrak{A}_a=\CA_1$ for all $a$.
  }
  
 In order to define the continuous sections of $\CA_{\infty}^{\mathrm (c)}$, we define, for $M\leq N$,    symmetrization maps $j_{NM}:  \CA^{\ot M}_1\raw  \CA^{\ot N}_1$ by
\beq
j_{NM}(A_M)=S_N(A_M\ot 1\ot\cdots \ot 1), \label{symmaps}
\eeq 
where one has $N-M$ copies of the unit $1\in\CA_1$ so as to obtain an element of $\CA_1^{\ot N}$. The symmetrization operator $S_N: \CA^{\ot N}_1\raw \CA^{\ot N}_1$ is given by (linear and continuous) extension of \beq
S_N(B_1\ot\cdots \ot B_N)=\frac{1}{N!}\sum_{\sg\in \GS_N} B_{\sg(1)}\ot\cdots\ot B_{\sg(N)}, \label{landc}
\eeq
where $\GS_N$ is the permutation group (i.e.\ symmetric group) on $N$ elements and $B_i\in\CA_1$ for all $i=1,\ldots,N$. For example, $j_{N1}:\CA_1\raw\CA_1^{\ot N}$ is given by
\beq
j_{N1}(B)=
\ovl{B}^{(N)}=\frac{1}{N}\sum_{k=1}^N 1\ot\cdots\ot B_{k}\ot 1\cdots \ot1,\label{B}\eeq
where $B_{k}$ is $B$ seen as an element of the $k$'th copy of $\CA_1$ in $\CA_1^{\ot N}$. 
In particular, for $\CA_1=M_2(\C)$ the frequency operator $f_N$ in $\CA^{\ot N}_1$ defined by \er{deffN}  is of this form, since from \er{abuse} we infer that
\beq f_N=j_{N1}(|1\ra\la 1|).\eeq
More generally, for $\CA_1=\CB(H)$ (the algebra of all bounded operators on a Hilbert space $H$), the operator that counts the frequency of the eigenstate
$|\lm\ra\in H$ of some observable $A$ upon $N$ measurements of $A$ is given by
\beq f_N=j_{N1}(|\lm\ra\la \lm|).\label{flm}\eeq
\begin{Definition}\label{defss}
We say that a sequence $(A_1,A_2,\cdots)$ with $A_N\in\CA_1^{\ot N}$
is \emph{symmetric} when  
\beq A_N=j_{NM}(A_M) \label{ass}
\eeq  for some fixed $M$ and all $N\geq M$. 
We call  $(A_1,A_2,\cdots)$ \emph{quasisymmetric} if for any $\varep>0$ there is an $N_{\varep}$ and a symmetric sequence $(A_1',A_2',\cdots)$ such that 
$\|A_N-A_N'\|< \varep$ for all $N\geq N_{\varep}$.
\end{Definition} 
Physically speaking,  the tail of a symmetric sequence entirely consists of `averaged' or `intensive' observables. which become macroscopic in the limit $N\raw\infty$.
Quasisymmetric sequences have the important property that they mutually commute in the limit $N\raw\infty$; more precisely, if  $(A_1,A_2,\cdots)$ and $(A_1',A_2',\cdots)$ are quasisymmetric sequences, then
\beq
\lni \| A_NA_N'-A_N'A_N\|=0. \label{aprc}
\eeq
Hence we see that in the limit $N\raw\infty$ the
quasisymmetric sequences organize themselves in a commutative \ca, which we call the
\ca\ of \emph{macroscopic observables} of the given large system. 
 To see that this limit algebra of macroscopic observables is isomorphic to $C(\CS(\CA_1))$, we complete the definition of the continuous field $\CA_{\infty}^{\mathrm (c)}$  
 by defining  its continuous sections.
  \begin{Theorem}
 For any unital \ca\ $\CA_1$, the fibers
 \begin{eqnarray}
 \CA_{N}^{\mathrm (c)} &=&  \CA_1^{\ot N};\nn \\
\CA_{\infty}^{\mathrm (c)} &=& C(\CS(\CA_1)) \label{fiberc2}
\end{eqnarray}
form a continuous field of \ca s over $\dot{\N}$ if the \ca\ $\CA^{\mathrm (c)}$ of  continuous sections is defined as follows:  a section 
\begin{eqnarray}
A&:& N\mapsto A_N\in \CA_1^{\ot N};\nn\\
 &:& \infty \mapsto A_{\infty}\in C(\CS(\CA_1))
\end{eqnarray}
 of the above field is declared to be continuous if 
 the sequence $(A_1,A_2,\cdots)$ is quasisymmetric, and
  \beq
 A_{\infty}(\om)=\lni\om^{\ot N}(A_N).\label{defAinf}\eeq 
\end{Theorem}
 Here $\om\in\CS(\CA_1)$ and $\om^{\ot N}\in \CS(\CA_1^{\ot N})$ is the tensor product of $N$ copies of $\om$, defined by (linear and continuous) extension of 
$ \om^{\ot N}(B_1\ot\cdots\ot B_N)=\om(B_1)\cdots\om(B_N)$; cf.\ \cite[Prop.\ 11.4.6]{KR2}.
 The limit \er{defAinf} then exists by definition of an approximately symmetric sequence:
if  $(A_1,A_2,\cdots)$  is symmetric  with \er{ass}, one has $\om^{\ot N}(A_N)=\om^{\ot M}(A_M)$ for $N>M$, so that the tail of the sequence $(\om^{\ot N}(A_N))$ is even independent of $N$. In the approximately symmetric case one easily proves that  $(\om^{\ot N}(A_N))$ is a Cauchy sequence.

{\bf Proof.} 
To prove that $\CA_{\infty}^{\mathrm (c)}$  is a continuous field, the main point is to show that
\beq \lni \| A_N\|=\|A_{\infty}\|, \label{normeq}
\eeq
  if $(A_1,A_2,\cdots)$ is quasisymmetric and $A_{\infty}$ is given by
\er{defAinf}. This is easy to show for symmetric sequences: assume \er{ass}, 
so that $\|A_N\|=\|j_{NN}(A_N)\|$  for $N\geq M$. By the $C^*$-axiom $\|A^*A\|=\|A^2\|$ it suffices to prove \er{normeq} for $A_{\infty}^*=A_{\infty}$, which implies $A_M^*=A_M$ and hence $A_N^*=A_N$ for all $N\geq M$. One then has $\| A_N\|=\sup\{|\rh(A_N)|, \rh\in\CS(\CA_1^{\ot N})\}$. 
Because of the special form of $A_N$ one may replace the supremum over 
the set $\CS(\CA_1^{\ot N})$ of all states on $\CA_1^{\ot N}$ by the supremum over the set $\CS^p(\CA_1^{\ot N})$ of all symmetric states (see Definition \ref{def5} below), which in turn
may be replaced by the supremum over the extreme boundary $\partial \CS^p(\CA_1^{\ot N})$ of $\CS^p(\CA_1^{\ot N})$. The latter consists of all states of the form $\rh=\om^{\ot N}$ \cite{Sto}, so that 
$\| A_N\|=\sup\, \{|\om^{\ot N}(A_N)|, \om\in\CS(\CA_1)\}$. This is actually equal
to $\| A_M\|=\sup\, \{|\om^{\ot M}(A_M)|\}$.  Now
the norm in $\CA_{\infty}^{\mathrm (c)}$ is $\|A_{\infty}\|=\sup\, \{|A_{\infty}(\om)|, \om\in\CS(\CA_1)\}$, and
by definition of $A_{\infty}$ one has $A_{\infty}(\om)=\om^{\ot M}(A_M)$. Hence
 \er{normeq} follows.

Given \er{normeq}, the theorem  follows from \cite[Prop.\ II.1.2.3]{book} and the fact that the set of functions $A_{\infty}$ on $\CS(\CA_1)$
arising in the said way are dense in $C(\CS(\CA_1))$ (equipped with the supremum-norm). This follows from the Stone--Weierstrass theorem, from which one infers that the functions in question exhaust $\CS(\CA_1)$.
\enp

We now turn to the continuous field  $\CA^{\mathrm (q)}$ defined by the quasilocal observables. 
\begin{Definition}
 A sequence $(A_1,A_2,\cdots)$ (where $A_N\in\CA_1^{\ot N}$, as before)  is called \emph{local} when for some fixed $M$ and all $N\geq M$ one has $A_N=A_M\ot 1\ot\cdots\ot 1$
 (where one has $N-M$ copies of the unit $1\in\CA_1$), and \emph{quasilocal} when  for any $\varep>0$ there is an $N_{\varep}$
and a local sequence  $(A_1',A_2',\cdots)$ such that 
$\|A_N-A_N'\|< \varep$ for all $N\geq N_{\varep}$. 
\end{Definition}

The inductive limit \ca\  $\underline{\lim}_{N}  \CA_1^{\ot N}$  then simply consists of all equivalence classes  $[A_1,A_2,\cdots]$ 
of quasilocal sequences $(A_1,A_2,\cdots)$  under the equivalence relation $(A_1,A_2,\cdots)\sim (B_1,B_2,\cdots)$ when $\lni\|A_N-B_N\|=0$. 
The  \ca ic structure on $\CA_{\infty}^{\mathrm (q)}$  is inherited from the quasilocal sequences
in the obvious (pointwise) way, except for  the norm, which is given by 
\beq \| [A_1,A_2,\cdots] \|=\lni \|A_N\|. \label{normql}\eeq
Each $\CA_1^{\ot N}$ is contained in $\CA_{\infty}^{\mathrm (q)}$ as a $C^*$-subalgebra by identifying $A_N\in \CA_1^{\ot N}$ with
the  equivalence class $[0,\cdots,0, A_N\ot 1,A_N\ot 1\ot1, \cdots]$ (where the zero's are irrelevant, of course; any entry could have been chosen). 
  \begin{Theorem}\label{Aqcf}
 For any unital \ca\ $\CA_1$, the fibers
 \begin{eqnarray}
 \CA_{N}^{\mathrm (q)} &=& \CA_1^{\ot N};\nn \\
\CA_{\infty}^{\mathrm (q)} &=&  \underline{\lim}_{N}  \CA_1^{\ot N} \label{fiberc4}
\end{eqnarray}
form a continuous field of \ca s over $\dot{\N}$ if the \ca\ $\CA^{\mathrm (q)}$ of  continuous sections is defined as follows:  a section 
\begin{eqnarray}
A&:& N\mapsto A_N\in \CA_1^{\ot N};\nn\\
 &:& \infty \mapsto A_{\infty}\in \underline{\lim}_{N}  \CA_1^{\ot N}
\end{eqnarray}
 of the above field is declared to be continuous if 
 the sequence $(A_1,A_2,\cdots)$ is quasilocal, and
  \beq
 A_{\infty}=[A_1,A_2,\cdots].\label{defAinfq}\eeq 
\end{Theorem}

{\bf Proof.} This time, the first property in Definition \ref{defcfca} is immediate from
\er{normql}. The other properties are either  trivial or follow from \cite[Prop.\ II.1.2.3]{book}.\enp
  \section{The Born rule}\label{sec4}
To see the relevance of the above considerations to the Born rule, we first rederive  \er{fop}.
For any $\CA_1$, we note that a state $\rh$ on  $\CA_1$ defines a state $\rh^{\ot\infty}$
on $\underline{\lim}_{N} \CA_N$ by
\beq \rh^{\ot\infty}([A_1,A_2,\ldots,])=\lni \rh^{\ot N}(A_N), \label{its}\eeq
which limit is easily seen to exist by first approximating a quasilocal sequence by a local one.

We take $\CA_1=B(H)$ and 
pick a unit vector  $|\ps\ra\in H$ with associated pure state $\ps$ on $\CA_1$. As in \er{its}, the infinite tensor product $\ps^{\ot\infty}$ defines a state on $\CA_{\infty}^{\mathrm (q)}$,
 which is pure by
\cite[Prop.\ 11.4.7]{KR2}. Hence the associated GNS-representation $\pi_{\ps^{\ot\infty}}(\CA_{\infty}^{\mathrm (q)})$ is irreducible; we may identify its carrier space
$H_{\ps^{\ot\infty}}$ with the separable component $H_{\ps}^{\ot\infty}$
of von Neumann's infinite tensor product $H^{\ot\infty}$ that contains $|\ps\ra^{\ot\infty}$
as the cyclic vector $\Om_{\ps^{\ot\infty}}$ of the GNS-construction (cf.\ \cite[Ch.\ 6]{EK} and also see Section \ref{sec2} above). 

Each operator $f_N$ defined by \er{flm} lies in $\CA_1^{\ot N}$
and hence in $\CA_{\infty}^{\mathrm (q)}$. Although $\lni f_N$ does not exist within 
 $\CA_{\infty}^{\mathrm (q)}$, 
one may consider a possible limit $\lni \pi_{\ps^{\ot\infty}}(f_N)$ as an operator on $H_{\ps^{\ot\infty}}$. This limit indeed exists in the strong operator topology, and commutes with 
all elements of $\pi_{\ps^{\ot\infty}}(\CA_{\infty}^{\mathrm (q)})$ (this is easily checked to be the case for any quasisymmetric sequence). Since $\pi_{\ps^{\ot\infty}}$ is irreducible, the limit operator must be a multiple of the unit, and using  \er{its} and \er{flm} one computes the constant as
\beq
\mathrm{s-}\lim_{N\raw\infty} \pi_{\ps^{\ot\infty}}(f_N)=|\la \lm|\ps\ra|^2\cdot 1.\label{Bb}\eeq
This generalizes \er{fop}, and also, to our mind,  gives an impeccable derivation of it. 
The type of derivation of the Born rule reviewed in Section \ref{sec2} is based on \er{Bb}, but
despite the fact that its mathematical status has now been clarified,  it faces the conceptual problems listed in that section. 

To solve these problems, we use the continuous field $\CA^{\mathrm (c)}$
instead of $\CA^{\mathrm (q)}$, again with $\CA_1=B(H)$.
 Identifying  a density matrix $\rh$ on $H$ with  a state on $B(H)$ in the usual way by
$\rh(B)=\mathrm{Tr}\, \rh B$, for a symmetric sequence with $A_N=j_{N1}(B)$ (see \er{B})
one easily finds
\beq A_{\infty}(\rh)=\mathrm{Tr}\, \rh B.\label{trace}\eeq
Our key application then  arises from the frequency operator \er{flm}, which amounts to the choice $B=|\lm\ra\la\lm|$. In that case
\er{trace} becomes
 \beq  f_{\infty}(\rh)=\la\lm| \rh|\lm\ra.\eeq
 In particular, if $|\ps\ra\in H$ is a unit vector and $\rh=|\ps\ra\la\ps|$, defining a vector state $\ps$ on $B(H)$  by $\ps(A)=\la\ps| A|\ps\ra$, one has
\beq f_{\infty}(\psi)=|\la \lm|\ps\ra|^2.\label{Bornrule}\eeq
This is the Born rule, at least formally.  To understand why this identification is
correct also conceptually, at least in the context of the Copenhagen Interpretation, one has to realize the following. Unlike its counterpart in \er{Bb}, the limit operator $f_{\infty}$ in  \er{Bornrule}  is \emph{by construction} an element of a commutative algebra, namely the \ca\ 
$C(\CS(\CA_1))$
of macroscopic observables attached to the $N$-fold duplication of $\CA_1$ for $N\raw\infty$.
According to Bohr's doctrine of classical concepts (cf.\ the Introduction), any statement about the quantum system described by $\CA_1$ has to be made through commutative \ca s 
$\mathcal{C}$ associated to $\CA_1$, and has to use ``the terminology of classical physics''.
This terminology includes the role of pure states as `truthmakers', in the sense that if $f:M\raw\R$ is a classical observable defined as a real-valued function on some phase space $M$, then
a point $\rh\in M$ validates the proposition $f=\lm$ for $\lm=f(\rh)$ with certainty. This is precisely what happens in  \er{Bornrule}, which uses $\mathcal{C}=C(\CS(\CA_1))$, hence
$M=\CS(\CA_1)$, and
states that in the classical state $\psi$, the observable $ f_{\infty}$ simply \emph{has} the (sharp) value $|\la \lm|\ps\ra|^2$. Thus one has a non-probabilistic statement in classical physics, which expresses a probabilistic observation about quantum physics.

The specific way in which $f_N$ converges to  $f_{\infty}$ as a continuous section
of $\CA^{\mathrm (c)}$, as well as its relationship to \er{Bb}, is clarified by 
the following device \cite{book,handbook}.
 \begin{Definition}
 A \emph{continuous field of states}
 on a continuous field of \ca s $(\CA,\{\CA_x\}_{x\in X}, \{\phv_x\}_{x\in X})$ over $X$  is a family of states $\om_x$ on $\CA_x$, defined for each $x\in X$, such that $x\mapsto \om_x(A_x)$ is continuous on $X$ for each $A\in \CA$ (i.e.\ for each continuous section 
 $x\mapsto \phv_x(A)\equiv A_x$ of the field of \ca s).
 \end{Definition}
   In the case at hand, where $X=\dot{\N}$, this only imposes the condition 
 \beq \om_{\infty}(A_{\infty})=\lni \om_N(A_N),\label{contf}\eeq
 for each continuous section $A$ of the field in question, which we take to be either 
 $\CA^{\mathrm (c)}$ or  $\CA^{\mathrm (q)}$. Indeed,  the relationship between these two continuous fields of \ca s is most easily studied through their respective continuous fields of states.
 
Any state $\om$  on $\CA_{\infty}^{\mathrm (q)}$ trivially  defines a  continuous fields of states on $\CA^{\mathrm (q)}$ by restriction, using the inclusion $\CA_1^{\ot N}\subset  \underline{\lim}_{N\raw\infty}  \CA_1^{\ot N}$ explained just above Theorem \ref{Aqcf}.
 The ensuing family of states $\om_N$ on $\CA_1^{\ot N}$
does not necessarily extend to a  continuous field on $\CA^{\mathrm (c)}$, and - especially in the context of the Born rule - it is interesting to find examples when they do.
\begin{Definition}\label{def5}
A state $\om$ on $\underline{\lim}_{N} \CA_1^{\ot N}$ is  \emph{symmetric} when each of its restrictions  to $\CA_1^{\ot N}$ is invariant under the natural action of the symmetric group $\GS_N$ on  $\CA_1^{\ot N}$ (under which $\sg\in\GS_N$ maps an elementary tensor $A_N=B_1\ot\cdots\ot B_N\in\CA_1^{\ot N}$ to $B_{\sg(1)}\ot\cdots\ot B_{\sg(N)}$). 
\end{Definition}
Such states were analyzed by St\o rmer \cite{Sto}, who proved a noncommutative version of  De Finetti's well-known \rep\ theorem in classical probability: any symmetric state
 $\om$ on $\underline{\lim}_{N} \CA_1^{\ot N}$ has a unique decomposition 
 \beq \om=\int_{\CS(\CA_1)} d\mu(\rh)\, \rh^{\ot\infty}, \label{Unn}
\eeq  where $\mu$ is a probability measure on $\CS(\CA_1)$, and $\rh^{\ot\infty}$ is defined as in \er{its}.

 \begin{Theorem}\label{propunn}
 Let  $\om$  be a symmetric state on  $\underline{\lim}_N \CA_1^{\ot N}$ with decomposition \er{Unn}, and let  $\om_N$ be the restriction of $\om$ to  $\CA_1^{\ot N}$. Define 
  a state  $\om_{\infty}$ on $\CA_{\infty}^{\mathrm (c)} = C(\CS(\CA_1))$
  by
 \beq \om_{\infty}(f)=\mu(f)\equiv \int_{\CS(\CA_1)} d\mu(\rh)\, f(\rh).\label{41}\eeq
 Then the family of states $\{\om_N, \om_{\infty}\}_{N\in\N}$ satisfies  \er{contf} for any 
 $A\in \CA^{\mathrm (c)}$  and hence defines a continuous
 family of states on $\CA^{\mathrm (c)}$.\end{Theorem}

{\bf Proof.} This is immediate from  \er{defAinf} and \er{Unn}.\footnote{Analogous results appear in the work of Unnerstall \cite{Unn1,Unn2}.
}\enp

We now see that the state $\om=\ps^{\ot\infty}$ used at the beginning of this section is an example of Definition \ref{def5}, for which the associated measure $\mu$ in \er{Unn} and \er{41} is the Dirac measure $\dl_{\ps}$ concentrated at $\ps\in \CS(\CA_1)$. The states $\om_N$
on $\CA_1^{\ot N}$ are, of course, given by $\om_N=\ps^{\ot N}$, and the function $N\mapsto
\om_N(f_N)$ has constant value $|\la \lm|\ps\ra|^2$. Hence one recovers the limit \er{Bornrule} either from  \er{contf} or from \er{41}, since $\dl_{\ps}(f_{\infty})=f_{\infty}(\ps)$; the fact that these computations coincide is an illustration of Theorem \ref{propunn}.

 %\newpage
\begin{small}
 \end{small}
\end{document}